\documentclass[superscriptaddress,showkeys,useAMS,twocolumn,nofootinbib]{revtex4-1}
\pdfoutput=1
\usepackage{graphics, graphicx, amsmath, amssymb}
\usepackage{epstopdf}
\usepackage{dsfont}
\usepackage{hyperref}
\usepackage{color}

\newcommand{\eq}[1]{\begin{equation}\begin{aligned}#1\end{aligned}\end{equation}}
\newcommand{\iu}{\text{i}}
\newcommand{\eu}{\text{e}}
\newcommand{\ha}{\hat{a}}
\newcommand{\had}{\hat{a}^\dagger}
\newcommand{\hb}{\hat{b}}
\newcommand{\hbd}{\hat{b}^\dagger}
\newcommand{\ket}[1]{\left|#1\right\rangle}
\newcommand{\bra}[1]{\left\langle#1\right|}

\newcommand{\expct}[1]{\left\langle#1\right\rangle}

\newcommand{\vac}{\ket{\text{vac}}}

\begin{document}
\title{Quantum-limited Euler angle measurements using anticoherent states}
\date{\today}
\author{Aaron Z. Goldberg}
\email{goldberg@physics.utoronto.ca}
\author{Daniel F. V. James}
\affiliation{Department of Physics, University of Toronto, Toronto, ON, M5S 1A7}
\begin{abstract}
Many protocols require precise rotation measurement. 
Here we present a general class of states that surpass the shot noise limit for measuring rotation around arbitrary axes. 
We then derive a quantum Cram\'er-Rao bound for simultaneously estimating all three parameters of a rotation (e.g., the Euler angles), and discuss states that achieve Heisenberg-limited sensitivities for all parameters; the bound is saturated by ``anticoherent" states [Zimba, Electron. J. Theor. Phys. 3, 143 (2006)] (we are reluctant to use ``anticoherent" to describe the states, but the name has become commonplace over the last decade). 
Anticoherent states have garnered much attention in recent years, and we elucidate a geometrical technique for finding new examples of such states.
Finally, we discuss the potential for divergences in multiparameter 
estimation due to singularities in spherical coordinate systems. 
 Our results are useful for a variety of quantum metrology and quantum communication applications.
\end{abstract}

\maketitle
\section{Introduction}

Estimating rotations is a highly relevant problem. Rotation measurements have applications in mathematics, physics, and beyond, ranging from geodesy \cite{Stocktonetal2011} and magnetrometry \cite{Wasilewskietal2010,Apellanizetal2018} to physiology \cite{Lametal2008} 
(see Ref. \cite{Szczykulskaetal2016} for a recent review). The problem of estimating rotations around a known axis is well-understood \cite{BondurantShapiro1984,Yurkeetal1986,SandersMilburn1995,Giovannettietal2004}, 
while rotation measurement around unknown axes is a relatively new endeavour \cite{Baganetal2001,PeresScudo2001,KolenderskiDemkowiczDobrzanski2008}. Measurement precision can be enhanced using special quantum input states in the known-axis case \cite{Caves1981,Xiaoetal1987}; here we investigate quantum enhancements for simultaneously estimating rotation angles and rotation axes.

Single parameter estimation has a long history \cite{MichelsonMorley1887}. One of the most famous examples is interferometry, in which the parameter in question is a phase imparted on a beam of light, which can be used to measure things such as biomolecules \cite{Huangetal1991,Tayloretal2013} and gravitational waves \cite{LIGO2011,Abbottetal2016}. 
Classical states of light are limited in their measurement precision by shot noise arising from photon statistics, leading to uncertainties bounded from below by $1/\sqrt{N}$, where $N$ is the number of photons involved in the measurement \cite{Dowling1998,Giovannettietal2004}. However, this is not a fundamental limit; cleverly designed schemes can take advantage of quantum correlations between photons to achieve the so-called Heisenberg limit, in which measurement uncertainties scale as $1/N$ \cite{Dowling1998,Giovannettietal2004,PezzeSmerzi2009,Toth2012}. In this work we seek similar quantum advantages for simultaneously estimating multiple parameters.

An arbitrary rotation in three dimensions is characterized by three parameters \cite{Grassia1998,GrafarendKuhnel2011}. These parameters can be the two angular coordinates of the rotation axis as well as the angle rotated around that axis, or any of the 12 triplets of Euler angles \cite{Diebel2006}. Here we seek to optimize estimates of the Euler angles; were the rotation axis to be known a priori, one could simply use single parameter estimation techniques. Nonetheless, any set of three parameters can be obtained from any other triplet.

Suitably designed quantum optical experiments can potentially enhance the simultaneous estimation of multiple parameters \cite{Humphreysetal2013}. For measuring phases imparted by either commuting or non-commuting operators, reductions in parameter uncertainties can be on the order of the number of parameters being estimated \cite{Humphreysetal2013,BaumgratzDatta2016}. A common technique for finding these enhancements involves the quantum Cram\'er-Rao bound, which bounds the covariance matrix between parameters being estimated by the inverse of the quantum Fisher information matrix (QFIM) \cite{BraunsteinCaves1994}. The quantum Cram\'er-Rao bound optimizes the covariance over all possible measurement techniques, and the QFIM depends on the chosen input state \cite{BraunsteinCaves1994,BRAUNSTEINetal1996}; therefore, an important task is finding quantum states that maximize the QFIM.

One area in which the QFIM has been studied is for reference frame alignment. Consider two parties who want to share some spatial information; to do so, they must know each other's coordinate system. Estimating the rotation required to align two coordinate systems has been studied, and it was found that ``anticoherent"
states maximize the QFIM \cite{KolenderskiDemkowiczDobrzanski2008}. This result is highly insightful for measurements of rotations about unknown axes.

{Anticoherent states are those whose polarization vectors vanish, and whose higher order polarization moments are isotropic \cite{Zimba2006}.} 
They are the furthest states from perfectly polarized states of light \cite{GoldbergJames2017}, with both classical and quantum notions of polarization vanishing for anticoherent states \cite{Zimba2006,Martinetal2010}.
{Because polarized light behaves more classically than unpolarized light, anticoherent states are in some sense the least classical quantum states \cite{Zimba2006,Aulbachetal2010,Hyllusetal2012,BaguetteMartin2017}.}
Anticoherent states have numerous mathematical and physical applications, relating to old problems of distributing points around a sphere \cite{Thomson1904,Whyte1952} and new challenges such as maximizing quantum entanglement \cite{Martinetal2010,Aulbachetal2010,Markham2011} or other notions of nonclassicality \cite{Giraudetal2010,Bjorketal2015}. %
Some of these states have already been created experimentally using light's orbital angular momentum degrees of freedom \cite{Bouchardetal2017}.
The states can be readily used for optimizing estimates of rotation parameters.

In this paper, we use the QFIM and the quantum Cram\'er-Rao bound to derive a bound on the covariance matrix for simultaneously estimating all three Euler angles. We find that anticoherent states can be used to achieve Heisenberg-limited variances in estimating all three rotation parameters, showing a quantum enhancement relative to estimation techniques using classical states. 
We also find that, regardless of parametrization, there exist angles for which the measurement precision diverges, and discuss a relation with the so-called ``hairy ball theorem" \cite{Milnor1978}. 
The multiparameter technique outperforms classical shot noise scaling everywhere, even in this diverging regime.
We then compare this scheme to other parametrizations of rotation parameters as well as combinations of single parameter estimation schemes, finding that we can always achieve quantum enhancements in measurement precision relative not just to the classical case but also to the best single parameter schemes. Finally, we provide an intuitive, alternative method for identifying new anticoherent states.

\section{Rotations and polarization}
\label{sec:rotations polarization}
We begin by considering polarization states of light; these are mathematically equivalent to any quantum states made from two harmonic oscillator modes. The two modes are associated with  operators $\ha$ and $\hb$ satisfying bosonic commutation relations $\left[\ha_i,\had_j\right]=\delta_{ij},\,\ha_i\in\left\{\ha,\hb\right\}$, such that a general state can be written as
\eq{\ket{\psi}=\sum_{m,n}c_{m,n}\ket{m,n},\quad \ket{m,n}\equiv \had\vphantom{a}^m\hbd\vphantom{a}^n\vac/\sqrt{m!n!}.} 
Using these operators we can define angular momentum operators \cite{Collett1970}
\eq{
	\hat{S}_0=\left(\had\ha+\hbd\hb\right)/2 \quad
	\hat{S}_x=\left(\had\hb+\hbd\ha\right)/2 \\
	 \hat{S}_y=-\iu\left(\had\hb-\hbd\ha\right)/2 \quad
	\hat{S}_z=\left(\had\ha-\hbd\hb\right)/2
	\label{eq:Quantum Stokes operators}
}
satisfying the usual $\mathfrak{su}$(2) algebraic equations
\eq{
	\left[\hat{S}_i,\hat{S}_j\right]=\iu\sum_{k=1}^3\epsilon_{ijk} \hat{S}_k\\
	\hat{S}_x^2+\hat{S}_y^2+\hat{S}_z^3=\hat{S}_0\left(\hat{S}_0+1\right).
	\label{eq:su2 algebra}
} Here, $\hat{S}_0$ is the total angular momentum operator, counting the total number of quanta in the system. 
 If, for example, $\ha$ and $\hb$ represent annihilation operators for two orthogonal polarizations of light, then we can associate the operators in Eq. (\ref{eq:Quantum Stokes operators}) with quantum Stokes operators, whose expectation values are the classical Stokes parameters (up to a normalization factor) \cite{JauchRohrlich1955,Collett1970}.
 
 The Stokes parameters contain all of the polarization information of classical states of light, denoted by the vectors $\mathbf{\hat{S}}\equiv\left(\hat{S}_x,\hat{S}_y,\hat{S}_z\right)$ and $\mathbf{S}=\expct{\mathbf{\hat{S}}}$ \cite{BornWolf1999}.
 The Stokes operators generate the SU(2) rotation operators 
 \eq{\hat{R}\left(\chi,\mathbf{n}\right)=\exp\left(\iu\chi\mathbf{\hat{S}}\cdot\mathbf{n}\right),\label{eq:R omega n}} 
 which rotate the Stokes vector $\mathbf{S}$ by angle $\chi$ about axis $\mathbf{n}=\left(\sin\theta\cos\phi,\sin\theta\sin\phi,\cos\theta\right)$. 
 The goal of this paper is to identify ways of measuring the three independent parameters of these rotation operators $\hat{R}$. 

The key concept of polarization is that it defines a preferred direction for a state.  A state is unpolarized if its Stokes vector is isotropic and therefore unchanged by rotations, which is only satisfied when $\mathbf{S}=\mathbf{0}$ \cite{BornWolf1999}. Quantum mechanically, however, higher-order moments are required to fully characterize a general state; there may still be some polarization information in these classically unpolarized states  \cite{PrakashChandra1971,Klyshko1992}. This prompted the definition of anticoherent states as those whose higher-order polarization moments are also unchanged under rotations. A ``$t$-anticoherent" state satisfies $\expct{\left(\mathbf{\hat{S}}\cdot\mathbf{n}\right)^k}=c_k$ for all positive integers $k\leq t$ and all unit vectors $\mathbf{n}$ \cite{Zimba2006}. States that are anticoherent to order $t$ have isotropic polarization moments up to order $t$, just as classically unpolarized states of light have polarization vectors that point nowhere.
The conditions for a state to be anticoherent have become more clear in recent years \cite{Crannetal2010,BannaiTagami2011,Baguetteetal2015,PereiraPaulPaddock2017}; in Section \ref{sec:2 anticoherent states}, we provide a geometrical formulation for finding new anticoherent states.

\section{Single parameter estimation}
\label{sec:single parameter estimation}
For a known axis $\mathbf{n}$, one can try to optimize measurements of the rotation angle $\chi$ around that axis by minimizing the variance in an estimate of $\chi$. One method of estimating $\chi$ is by measuring the projection $\hat{P}=\ket{\psi}\bra{\psi}$ of an initially prepared pure state $\ket{\psi}$ onto the rotated state $\hat{R}\ket{\psi}$:
$\expct{\hat{P}}=\left|\bra{\psi}\hat{R}\left(\chi,\mathbf{n}\right)\ket{\psi}\right|^2$,
where expectation values are henceforth taken with respect to the rotated state. For small angles $\chi$, one can expand the exponential in Eq. (\ref{eq:R omega n}) to find $\expct{\hat{P}}=1-\chi^2\text{Var}\left[\mathbf{\hat{S}}\cdot\mathbf{n}\right]+\mathcal{O}\left(\chi^4\right)$, for variances $\text{Var}\left[X\right]=\expct{X^2}-\expct{X}^2$. This can be used to calculated the variance of the estimated angle
\eq{
\text{Var}\left[\chi\right]=\frac{\text{Var}\left[\hat{P}\right]}{\left|\frac{\partial\expct{\hat{P}}}{\partial\chi}\right|^2}\approx \frac{1}{4\text{Var}\left[\mathbf{\hat{S}}\cdot\mathbf{n}\right]}.
}

Coherent state inputs with average photon number $N$, such as $\eu^{-N/2}\sum_{m=0}^{\infty}\frac{N^{m/2}}{\sqrt{m!}}\ket{m,0}$, have $4\text{Var}\left[\hat{\mathbf{S}}\cdot\mathbf{n}\right]=N$. These classical states can at best achieve the shot noise precision $\Delta\chi\equiv\sqrt{\text{Var}\left[\chi\right]}=1/\sqrt{N}$. In comparison,
the NOON states $\frac{\ket{N,0}+\ket{0,N}}{\sqrt{2}}$ satisfy $4\text{Var}\left[\mathbf{\hat{S}}\cdot\mathbf{n}\right]=N^2\cos^2\theta+N\sin^2\theta$, and so can achieve Heisenberg-limited precisions $\Delta \chi=1/N$ for rotations around a unit vector $\mathbf{n}$ aligned with $\theta=0$ axis. This is an important example of the fact that particular input states can provide quantum-enhanced sensitivities in parameter estimation.

If we do not specify a measurement scheme, the quantum Cram\'er-Rao bound tells us that $\Delta\chi\geq1/\sqrt{I}$, for quantum Fisher information \cite{SahotaQuesada2015} \eq{I=4\left[\bra{\psi}\left(\partial_{\chi}\hat{R}\right)^\dagger\partial_{\chi}\hat{R}\ket{\psi} - \left|\bra{\psi}\hat{R}^\dagger\partial_{\chi}\hat{R}\ket{\psi}\right|^2\right].}
The measurement scheme of projecting onto the initial state, described above, saturates the Cram\'er-Rao bound in the small-angle limit, which is always possible for single parameter estimation \cite{Hyllusetal2012,Humphreysetal2013,Szczykulskaetal2016}. The Fisher information together with the quantum Cram\'er-Rao bound can thus be used as a way of determining input states that will achieve optimal sensitivities for a particular transformation.

It is clear that states with isotropic $\text{Var}\left[\mathbf{\hat{S}}\cdot\mathbf{n}\right]=\mathcal{O}(N^2)$ are useful for estimating rotations about arbitrary, known axes $\mathbf{n}$. These states will achieve Heisenberg-scaling precisions regardless of the rotation axis $\mathbf{n}$, albeit with slightly less sensitivity than the NOON states rotating about the $\theta=0$ axis. The 2-anticoherent states have $\expct{\mathbf{\hat{S}}}=\mathbf{0}$ and $\expct{\left(\mathbf{\hat{S}}\cdot\mathbf{n}\right)^2}=\expct{\hat{S_0}\left(\hat{S_0}+1\right)}/3$, which allows the variances to scale quadratically with the number of quanta in the initial states. Recent experiments have used these states to achieve Heisenberg-scaling sensitivities for rotations around a variety of rotation axes \cite{Bouchardetal2017}. We presently investigate these states in the context of measuring rotations whose axes are not known a priori.

\section{Multiparameter estimation}
In this section we investigate the most sensitive techniques for simultaneously measuring changes in all three Euler angles of a rotation.
\label{sec:multiparameter estimation}
\subsection{Quantum Fisher information matrix for Euler angle estimation}
The QFIM for pure states has components \cite{Paris2009}
\eq{    \left[\mathbf{I}_{\boldsymbol{\theta}}\right]_{l,m}=\frac{1}{2}
	\bra{\psi_{\boldsymbol{\theta}}}
	L_lL_m+L_mL_l
	\ket{\psi_{\boldsymbol{\theta}}}
}
for symmetric logarithmic derivatives
\eq{
	L_i=2\left(\ket{\partial_{\theta_i}\psi_{\boldsymbol{\theta}}}\bra{\psi_{\boldsymbol{\theta}}}
	+
	\ket{\psi_{\boldsymbol{\theta}}}\bra{\partial_{\theta_i}\psi_{\boldsymbol{\theta}}}
	\right).
}
Here, $\ket{\psi_{\boldsymbol{\theta}}}=\hat{R}\left(\boldsymbol{\theta}\right)\ket{\psi_0}$, for input state $\ket{\psi_0}$ and some triplet of rotation parameters $\boldsymbol{\theta}$. The operator $L_\chi$ is relatively easy to compute given the parametrization $\hat{R}=\exp\left(\iu\chi\mathbf{\hat{S}}\cdot\mathbf{n}\right)$, because in taking the derivative $\partial_\chi\hat{R}$ one need not consider operator ordering. The other components can be calculated using Ref. \cite{Wilcox1967}'s more involved technique; for clarity, we concentrate on a simpler Euler angle parametrization for our subsequent discussion to avoid introducing this new technique.

Following the notation in Ref. \cite{Camposetal1989} discussing the SU(2) representation of beam splitters, we parametrize our rotation operators by
\eq{
	\hat{R}\left(\Phi,\Theta,\Psi\right)=\eu^{-\iu\Phi \hat{S}_z}\eu^{-\iu\Theta \hat{S}_y}\eu^{-\iu\Psi \hat{S}_z}.
}
Our goal is to estimate the parameters $\left(\Phi,\Theta,\Psi\right)$.
For this we must evaluate the symmetric logarithmic derivatives defining the quantum Fisher information, which rely on derivatives of $\hat{R}$ with respect to each of the three angles:
$\partial_{\theta_i}\hat{R}=-\iu\hat{H}_{{\theta}_i}\hat{R},\, {\theta}_i\in\left(\Phi,\Theta,\Psi\right)$, for operators
\eq{
	&\hat{H}_\Phi\equiv \hat{S}_z\\
	&\hat{H}_\Theta\equiv \eu^{-\iu\Phi\hat{S}_z}\hat{S}_y\eu^{\iu\Phi\hat{S}_z}=-\sin\Phi\hat{S}_x+\cos\Phi\hat{S}_y\\
	&\hat{H}_\Psi\equiv\hat{R}\hat{S}_z\hat{R}^\dagger=\sin\Theta\cos\Phi\hat{S}_x+\sin\Theta\sin\Phi\hat{S}_y+\cos\Theta\hat{S}_z .
}

Then we find that $\left[\mathbf{I}_{\boldsymbol{\theta}}\right]_{l,m}$ takes the form
\eq{
	\left[\mathbf{I}_{\boldsymbol{\theta}}\right]_{l,m}=4\text{Cov}\left\{\hat{H}_l,\hat{H}_m\right\},
}
where we 
use $\text{Cov}\left\{X,Y\right\}=\expct{\frac{XY+YX}{2}}-\expct{X}\expct{Y}$ and expectation value are taken with respect to the rotated state $\ket{\psi_{\boldsymbol{\theta}}}$. All that remains is to find states $\ket{\psi_{\boldsymbol{\theta}}}$ that maximize the amount of information in this matrix.

\subsection{Quantum Fisher information matrix for optimum input states}

We assume that the input states have exactly $N$ quanta, as these will always perform at least as well as superposition states with various numbers of quanta \cite{KolenderskiDemkowiczDobrzanski2008}. It was shown that the best states for estimating the three components of a reference frame have $\expct{\mathbf{\hat{S}}}=\mathbf{0}$ and $\expct{\hat{S}_i\hat{S}_j}=\delta_{ij}\frac{N}{2}\left(\frac{N}{2}+1\right)/3$ \cite{KolenderskiDemkowiczDobrzanski2008}; these are the 2-anticoherent states \cite{Zimba2006}.
 We calculate the QFIM for 2-anticoherent states with $N$ quanta, in the $(\Phi,\Theta,\Psi)$ basis:
\eq{
	\mathbf{I}_{\boldsymbol{\theta}}=\frac{N\left(N+2\right)}{3}\begin{pmatrix}
		1&0&\cos\Theta\\0&1&0\\\cos\Theta&0&1	
	\end{pmatrix}.\label{eq:QFIM}
}

The quantum Cram\'er-Rao bound says that the covariance matrix for the parameters $(\Phi,\Theta,\Psi)$ satisfies the inequality \eq{\text{Cov}\left\{\boldsymbol{\theta}\right\}\geq \mathbf{I}_{\boldsymbol{\theta}}^{-1}
	=\frac{3}{N\left(N+2\right)}\begin{pmatrix}
		\frac{1}{\sin^2\Theta}&0&-\frac{\cos\Theta}{\sin^2\Theta}\\0&1&0\\-\frac{\cos\Theta}{\sin^2\Theta}&0&\frac{1}{\sin^2\Theta}	
	\end{pmatrix}
	;\\
} this bound can be saturated for pure states	with real symmetric logarithmic derivatives \cite{Hyllusetal2012,Humphreysetal2013,Szczykulskaetal2016}, which is always the case here. 

This result
gives excellent scaling with $N$. The actual measurement precisions have very small bounds for $\sin\Theta\approx 1$, but are worse when $\sin\Theta\approx 0$, which can be expected from the chosen Euler angle parametrization. This is because, for $\Theta=0$, the beam splitter simply acts as $\hat{R}=\eu^{-\iu\left(\Phi+\Psi\right)\hat{S}_z}$, and so one would only ever be able to estimate $\Theta$ and the sum $\Phi+\Psi$. Alternatively, we can see this divergence by considering the difference in parameters $\Phi-\Psi$. The eigenvector $\left(1,0,-1\right)$ of the QFIM given in Eq. (\ref{eq:QFIM}) corresponding to the difference $\Phi-\Psi$ has eigenvalue proportional to $1-\cos\Theta$, which vanishes at $\Theta=0$ (see Ref. \cite{WittenburgLilov2003} for further discussion). Our estimation scheme does well everywhere other than at this angle.

For any $\Theta >0$ we thus get Heisenberg scaling in the variance of estimating $\Phi$, $\Theta$, and $\Psi$ simultaneously:
\eq{
	\text{Var}\left[\Phi\right]+\text{Var}\left[\Theta\right]+\text{Var}\left[\Psi\right]\geq \text{Tr}\left[\mathbf{I}^{-1}_{\boldsymbol{\theta}}\right]\\=\frac{3}{N(N+2)}\left(1+\frac{2}{\sin^2\Theta}\right).
\label{eq:fisher inverse trace}
}
Eq. (\ref{eq:fisher inverse trace}) shows that quantum enhancements can be achieved in the simultaneous estimation of all three rotation parameters.

\subsection{Comparison to single parameter estimation and other rotation parametrizations}
We can compare our result in Eq. (\ref{eq:fisher inverse trace}) to the best possible single parameter estimation techniques, as well as parametrizations other than $\left(\Phi,\Theta,\Psi\right)$. 
The optimum single parameter estimation techniques use NOON states. A scheme that uses three NOON states to measure rotations around three arbitrary axes, with $N/3$ particles in each state, only yields the quantum Cram\'er-Rao bound
(see Appendix \ref{app:single parameter variances})
\eq{
	\text{Var}\left[\Phi\right]+\text{Var}\left[\Theta\right]+\text{Var}\left[\Psi\right]\geq 
	\left(\frac{3}{N}\right)^2\\
	\times\left(1+\frac{1}{\cos^2\theta_1+\frac{1}{N}\sin^2\theta_1}+\frac{1}{\cos^2\theta_2+\frac{1}{N}\sin^2\theta_2}\right),
	\label{eq:single parameter variances}
}
where $\theta_1$ is the angle between one of the chosen axes and $\left(-\sin\Phi,\cos\Phi,0\right)$, and $\theta_2$ is the angle between another chosen axis and $\left(\sin\Theta\cos\Phi,\sin\Theta\sin\Phi,\cos\Theta\right)$.

Even for the most fortuitous choice of axes ($\theta_1=\theta_2=0$ versus $\Theta=\pi/2$),
our multiparameter scheme outperforms this single parameter scheme by a factor $d+2d/N$, where $d=3$ is the number of parameters being estimated. This is because one can only use $N/d$ quanta per measurement in the single parameter scheme, which is similar to the $\mathcal{O}(d)$ enhancements found by Ref. \cite{Humphreysetal2013,BaumgratzDatta2016} in using $d$-mode schemes to simultaneously estimate $d$ parameters. 
However, one should not combine the Cram\'er-Rao bounds for single parameters when multiple parameters are unknown. The off-diagonal elements of the QFIM are what determine its singularities; it is unfair to compare to single parameter estimation schemes when none of the parameters necessary to determine the optimal measurement parameters are known a priori (e.g., $\theta_1$ and $\theta_2$). 
The multiparameter technique outlined above is thus necessary for simultaneously estimating $\Phi$, $\Theta$, and $\Psi$, always beating single parameter schemes.

We further note that every parametrization of the rotation parameters has divergences in the trace of the covariance matrix for particular angles. For example, if we use the rotation operators $\hat{R}\left(\alpha,\beta,\gamma\right)=\eu^{-\iu \alpha\hat{S}_x}\eu^{-\iu \beta\hat{S}_y}\eu^{-\iu \gamma\hat{S}_z}$ in our multiparameter scheme, 
we find that the trace of the covariance matrix again achieves Heisenberg scaling, but with divergences at angles satisfying $\sin\left(2\alpha\right)=\cos\left(\beta\right)/\sin^2\left(\beta/2\right)$.
Similarly, parametrizing a rotation by its rotation angle and its rotation axis, as in Eq. (\ref{eq:R omega n}), achieves Heisenberg scaling with divergences at $\chi=0$ and $\theta=0$ (see Appendix \ref{app:divergences} for further discussion of such coordinate singularities). 

Coordinate singularities in three-dimensional parametrizations of rotations relate to the assertion of Brouwer's fixed-point theorem, the so-called ``hairy-ball theorem," 
that nonvanishing continuous  tangent vector fields on the sphere $S^2$ do not exist \cite{Milnor1978}. The theorem implies that a function mapping the eigenvector unchanged by a rotation matrix to its orthogonal basis vectors cannot be continuous, which forces three-dimensional parametrizations of $S^2$ to always be singular somewhere \cite{GrafarendKuhnel2011}. 
The QFIM being singular is a signature of the singularities present in every choice of three-dimensional parametrizations of $S^2$ \cite{WittenburgLilov2003}.

\section{Finding 2-anticoherent states}
\label{sec:2 anticoherent states}

The states that optimize the QFIM for estimating Euler angles are 2-anticoherent states: classically unpolarized states with isotropic variances in their Stokes operators. 
The original requirement for an $N$-qubit state 
$\ket{\psi^{(N)}}=\sum_{m=0}^{N}c_m\ket{m,N-m}$
to be $2$-anticoherent can be written as
\eq{\mathbf{S}=\mathbf{0},\quad \mathcal{S}=\frac{N(N+2)}{12}\mathds{I},
\label{eq:2anticoherent requirements}}
where we define the Hermitian tensor $\mathcal{S}$ with components $\mathcal{S}_{i,j}=\expct{\hat{S}_i\hat{S}_j}$ \cite{Zimba2006}. 

To identify such states, the usual approach is via the Majorana representation \cite{Majorana1932}. The Majorana representation allows us to uniquely write $N$-qubit states as $\ket{\psi^{(N)}}\propto\prod_{k=1}^{N}\had_{\theta_k,\phi_k}\ket{\text{vac}}$. Then, the $N$ creation operators $\had_{\theta_k,\phi_k}=\cos\frac{\theta_k}{2}\had+\eu^{\iu\phi_k}\sin\frac{\theta_k}{2}\hbd$ uniquely map the state $\ket{\psi^{(N)}}$ to the $N$ points $\left\{\left(\theta_m,\phi_m\right)\right\}$ on the unit sphere, known as the Poincar\'e sphere in the context of polarization
\cite{Migdaletal2014,Bjorketal2015}.

A deep conjecture relating the Majorana representation of anticoherent states to spherical designs was proposed in 2010 \cite{Crannetal2010}, but counterexamples were elucidated shortly thereafter \cite{BannaiTagami2011}. A fruitful new approach for numerically finding these states has recently come to light \cite{Giraudetal2015,Baguetteetal2015,BaguetteMartin2017}; we comment briefly on earlier approaches and 
show an elegant, geometrical method for finding 2-anticoherent states that has not yet been elucidated.
\subsection{Mathematical scheme}
Similar to the method in Ref. \cite{Baguetteetal2015}, we present simple mathematical criteria for finding 2-anticoherent states. 
All 2-anticoherent states must have
\eq{\sum_{m=0}^{N}\left|c_m\right|^2m=\frac{N}{2},\quad \sum_{m=0}^{N}\left|c_m\right|^2m^2=\frac{N(2N+1)}{6},} due to $\expct{\hat{S}_z}=0$ and $\expct{\hat{S}_z^2}=N\left(N+2\right)/12$, in addition to the usual normalization $\sum_m\left|c_m\right|^2=1$. The other conditions can all be satisfied if we impose the additional requirement that $c_mc_{m+1}=c_mc_{m+2}=0$ for all $m$ (i.e., the spacing between each nonzero $c_m$ should be at least 2 values of $m$). This yields a set of three equations for the real parameters $\left|c_m\right|^2$, which can be solved analytically or numerically for a given choice of nonzero $\left\{c_m\right\}$ (see especially Ref. \cite{Baguetteetal2015} for interesting numerical results).

As an example, we give an analytical solution for systems with four nonzero coefficients $c_{N/4}$, $c_{N/2}$, $c_{3N/4}$, and $c_{N}$ (we choose this example because there are no 2-anticoherent states listed in Refs. \cite{KolenderskiDemkowiczDobrzanski2008,Baguetteetal2015} with exactly four nonzero coefficients). We find the infinite family of states
\begin{widetext}
	\eq{
		\ket{\psi_4^{(N)}}=c\ket{N,0}+\eu^{\iu\phi_1}\sqrt{\frac{2(2+N)}{3N}-3c^2}\ket{\frac{3N}{4},\frac{N}{4}}
		+\eu^{\iu\phi_2}\sqrt{3c^2-\frac{8+N}{3N}}\ket{\frac{N}{2},\frac{N}{2}}
		+\eu^{\iu\phi_3}\sqrt{\frac{2(2+N)}{3N}-c^2}\ket{\frac{N}{4},\frac{3N}{4}},
	}
\end{widetext}
for arbitrary $c\in\left(\frac{8+N}{9N},\frac{4+2N}{9N}\right)$ and $N\geq 12$. The states $\ket{\psi_4^{(N)}}$, and other easy-to-find 2-anticoherent states, can thus be used to achieve Heisenberg scaling of $\mathcal{O}(1/N)$ in the precision of simultaneously estimating all three Euler angles of a rotation.

\subsection{Geometrical scheme}
\label{sec:geometrical scheme}
There are important geometrical properties of 2-anticoherent states that can be used to find new such states without solving systems of linear equations. These make use of the Majorana representation.
To optimize estimates of the Euler angles, one seeks states that are highly sensitive to rotations. These optimal states, the 2-anticoherent states, are found to be states with highly symmetric Majorana representations \cite{KolenderskiDemkowiczDobrzanski2008,Bjorketal2015}.

\subsubsection{Platonic solids and the Majorana representation}
The problem of distributing points symmetrically about a sphere is not new. It has been studied in relation to mathematics \cite{Whyte1952}, biology \cite{Tammes1930}, and quantum entanglement \cite{Martinetal2010}. One of the earliest results for distributing points symmetrically uses the Platonic solids; the vertices of any of the five Platonic solids will be symmetrically spaced about a sphere circumscribing the solid.

States whose Majorana representations form a Platonic solid are always anticoherent to order 2 or higher \cite{Zimba2006,KolenderskiDemkowiczDobrzanski2008}. For example, the  state
\eq{\ket{\psi}&=\frac{1}{\sqrt{3}}\ket{4,0}+\sqrt{\frac{2}{3}}\ket{1,3}=\left(\frac{\had\vphantom{a}^4}{\sqrt{72}}+\frac{\had\hbd\vphantom{a}^3}{3}\right)\ket{\text{vac}},}
which is the most nonclassical state with $N=4$, a maximally entangled state of four qubits, and the most sensitive $N=4$ state for reference frame alignment, has vertices that form a tetrahedron, one of the Platonic solids \cite{Hyllusetal2012}. Moreover, states with $m$ Majorana points at each of the vertices of a Platonic solid, like
\eq{
	\ket{\psi}&\propto\left(\frac{\had\vphantom{a}^4}{\sqrt{72}}+\frac{\had\hbd\vphantom{a}^3}{3}\right)^m\ket{\text{vac}},
}
are also anticoherent.\footnote{This little-known fact is mentioned in a footnote in Ref. \cite{Zimba2006} without proof; we show why this is true in the next section.
} 
Platonic solids with $m$ Majorana points at each vertex can thus be used to increase the particle number $N$ in the Heisenberg scaling for estimating the Euler angles of a rotation (Fig. \ref{fig:tetrahedrons}).

\subsubsection{States with two rotational symmetries}
\label{sec:two rotational symmetries}

We here present a method for finding new 2-anticoherent states with arbitrarily large particle number $N$. We show that
any state with at least two discrete rotational symmetries along independent axes $\ket{\psi}=\hat{R}_1\ket{\psi}=\hat{R}_2\ket{\psi}$ will satisfy this property, and use the Majorana representation to find states with this property. 

The key to this result is the fact that the quantities $\mathbf{S}$ and $\mathcal{S}$ transform as vectors and tensors, respectively, under the transformation $\ket{\psi}\to\hat{R}\ket{\psi}$. For a rotation $\hat{R}\left(\chi,\mathbf{n}\right)$, we have $\mathbf{S}\to \mathcal{R}\mathbf{S}$ and $\mathcal{S}\to \mathcal{R}\mathcal{S}\mathcal{R}^{-1}$, where the rotation matrix $\mathcal{R}$ is given by the famous Rodrigues rotation formula
\begin{widetext}
	\eq{
		\mathcal{R}=\begin{pmatrix}
			\cos \chi +n_x^2 \left(1-\cos \chi\right) & n_x n_y \left(1-\cos \chi\right) + n_z \sin \chi & n_x n_z \left(1-\cos \chi\right) - n_y \sin \chi \\ 
			n_y n_x \left(1-\cos \chi\right) - n_z \sin \chi & \cos \chi + n_y^2\left(1-\cos \chi\right) & n_y n_z \left(1-\cos \chi\right) + n_x \sin \chi \\ 
			n_z n_x \left(1-\cos \chi\right) + n_y \sin \chi & n_z n_y \left(1-\cos \chi\right) - n_x \sin \chi & \cos \chi + n_z^2\left(1-\cos \chi\right)
		\end{pmatrix}.
	}
\end{widetext}
(see Ref. \cite{Leonhardt2003} for a list of the rotation matrices generated by the Stokes operators). If a state is unchanged via a rotation $\hat{R}$, the corresponding transformation must yield $\mathbf{S}=\mathcal{R}\mathbf{S}$ and $\mathcal{S}=\mathcal{R}\mathcal{S}\mathcal{R}^{-1}$.

A vector that is unchanged by two rotations about independent axes must have zero length. To see this, consider a vector unchanged by a rotation around a single axis; the vector must point along the axis of rotation. No vector can point along two independent axes without being the zero vector. Therefore, all states with the property 
\eq{\ket{\psi}=\hat{R}(\chi_1,\mathbf{n}_1)\ket{\psi}=\hat{R}\left(\chi_2,\mathbf{n}_2\right)\ket{\psi},\quad \mathbf{n}_1\nparallel \mathbf{n}_2 \label{eq:two symmetries}} must be at least 1-anticoherent, with $\mathbf{S}=\mathbf{0}$.

For the property given by Eq. (\ref{eq:two symmetries}) to guarantee that $\mathcal{S}
\propto\mathds{I}$, we must further require that $\chi_1,\chi_2\neq\pi$; i.e., neither $\mathcal{R}_1$ nor $\mathcal{R}_2$ can be reflection matrices. Without this condition, states such as NOON states would satisfy the property Eq. (\ref{eq:two symmetries}), but these are not 2-anticoherent states. 

Eq. (\ref{eq:two symmetries}) guarantees that the eigenvectors of $\mathcal{S}$ are all degenerate. The degeneracy follows from that fact that $\mathcal{S}$ commutes with both $\mathcal{R}_1$ and $\mathcal{R}_2$, because $\mathcal{S}=\mathcal{R}_1\mathcal{S}\mathcal{R}_1^{-1}=\mathcal{R}_2\mathcal{S}\mathcal{R}_2^{-1}$. The eigenvalue equation $\mathcal{S}\mathbf{v}=\lambda \mathbf{v}$ then guarantees that $\mathcal{R}_1\mathbf{v}$ and $\mathcal{R}_2\mathbf{v}$ are also eigenvectors of $\mathcal{S}$ with the same eigenvalue $\lambda$. Of note, the only real eigenvectors of $\mathcal{R}_1$ and $\mathcal{R}_2$ are proportional to $\mathbf{n}_1$ and $\mathbf{n}_2$, respectively, because rotations in three dimensions only have a single axis that they leave unchanged. We can then always find one of the three linearly independent eigenvectors of the $3\times 3$ Hermitian tensor $\mathcal{S}$ that is outside of the span of $\mathbf{n}_1$ and $\mathbf{n}_2$, call it $\mathbf{v}_0$, and guarantee that $\mathbf{v}_0$, $\mathcal{R}_1\mathbf{v}_0$, and $\mathcal{R}_2\mathbf{v}_0$ span 3 dimensions. The entire eigenspace of $\mathcal{S}$ is thus spanned by vectors with degenerate eigenvalues;
this implies that $\mathcal{S}$ is proportional to the identity matrix.

The proportionality constant in $\mathcal{S}\propto\mathds{I}$ is fixed by Eq. (\ref{eq:su2 algebra}) to yield the 2-anticoherence properties given in Eq. (\ref{eq:2anticoherent requirements}):
\eq{\mathbf{S}=\mathbf{0},\quad\mathcal{S}=\frac{N(N+2)}{12}\mathds{I}.}
Any state with two independent rotational symmetries [Eq. (\ref{eq:two symmetries})] will achieve Heisenberg scaling in estimating rotation parameters.

\subsubsection{Applications of the geometrical condition}
\begin{figure}
	\includegraphics[width=\columnwidth]
	{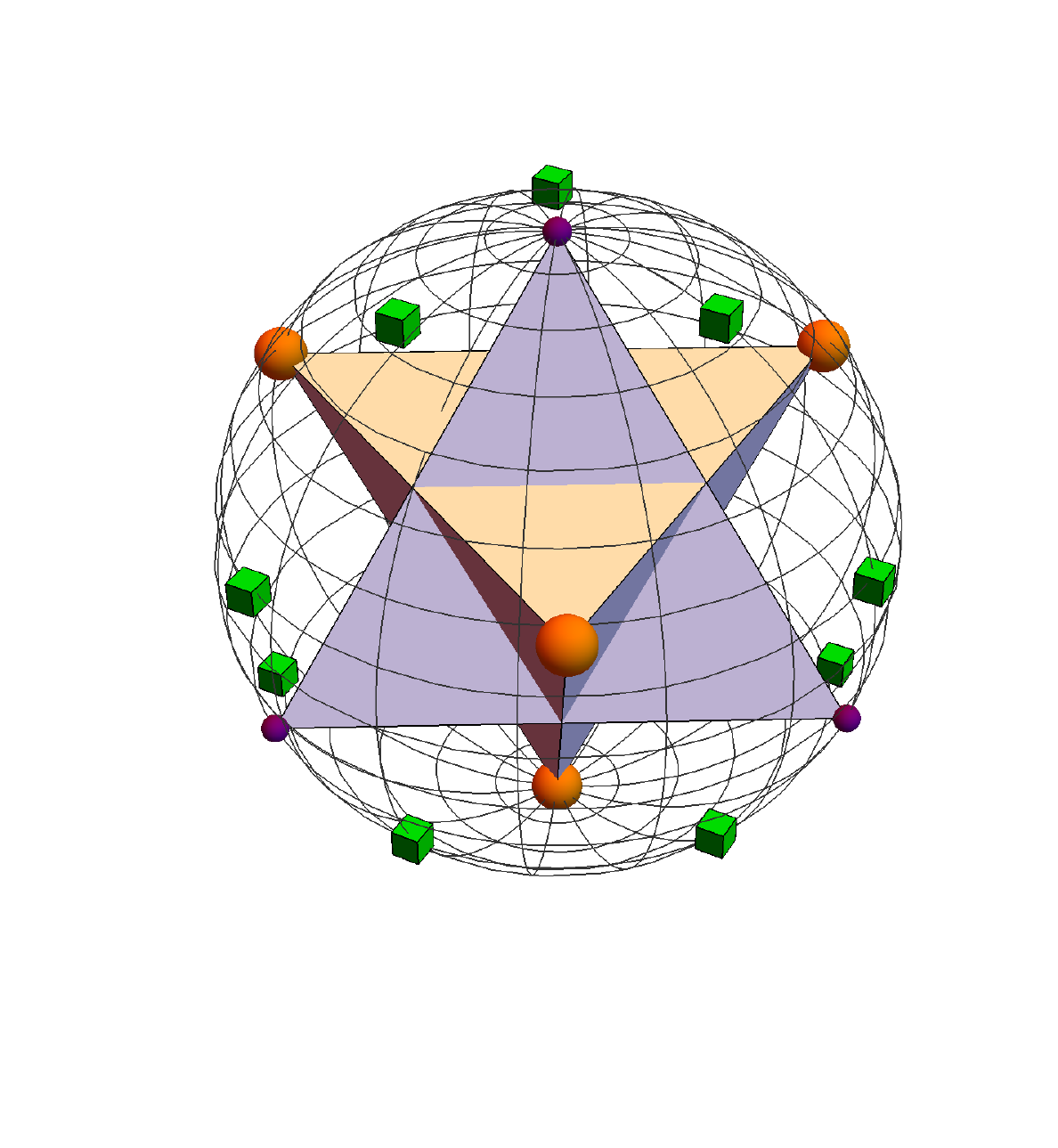}	\caption{Example of the Majorana representation of a quantum state that can be used to achieve Heisenberg-limited sensitivities $\mathcal{O}\left(1/N\right)$ for simultaneously estimating all three parameters of a rotation. The image shows two intersecting tetrahedra that are duals to each other, one coloured in purple and the other in orange. The vertices of the respective tetrahedra are on the surface of the sphere, coloured as purple (small) and orange (large) points. Also shown as green cubes are the vertices of a truncated tetrahedron aligned with the purple tetrahedron. A state whose Majorana representation has $m$ degenerate points at each of the (purple, small, spherical) vertices of one tetrahedron, $n$ degenerate points at the (orange, large, spherical) vertices of the other tetrahedron, and $k$ degenerate points at the (green, cubic) vertices of the truncated tetrahedron is a 2-anticoherent state, with $N=4(m+n+3k)$.}
	\label{fig:tetrahedrons}
\end{figure}

All of the Platonic solids have multiple discrete rotational symmetries, about axes defined by the lines from the centre of the sphere through any of the solids' vertices or through the middle of any of the solids' faces. These symmetries are a property of the geometry alone, and so states whose Majorana representations have $m$-fold degeneracies at each of the vertices of a Platonic solid are 2-anticoherent states.

Moreover, the duals of a polyhedron share its rotational symmetries, so any state whose Majorana representation features $m$-fold degeneracies at the vertices of a Platonic solid as well as $n$-fold degeneracies at the vertices of the Platonic solid's dual is also 2-anticoherent. For example, a state whose Majorana constellation has $m$ points at each of the vertices of a cube and $n$ points at each of the vertices of the cube's dual, an octahedron, is a 2-anticoherent state (see Fig. \ref{fig:tetrahedrons} for a similar example). Combinations of Platonic solids and their duals can be used to measure rotation parameters with high sensitivity.

\begin{figure}
	\includegraphics[scale=0.95]{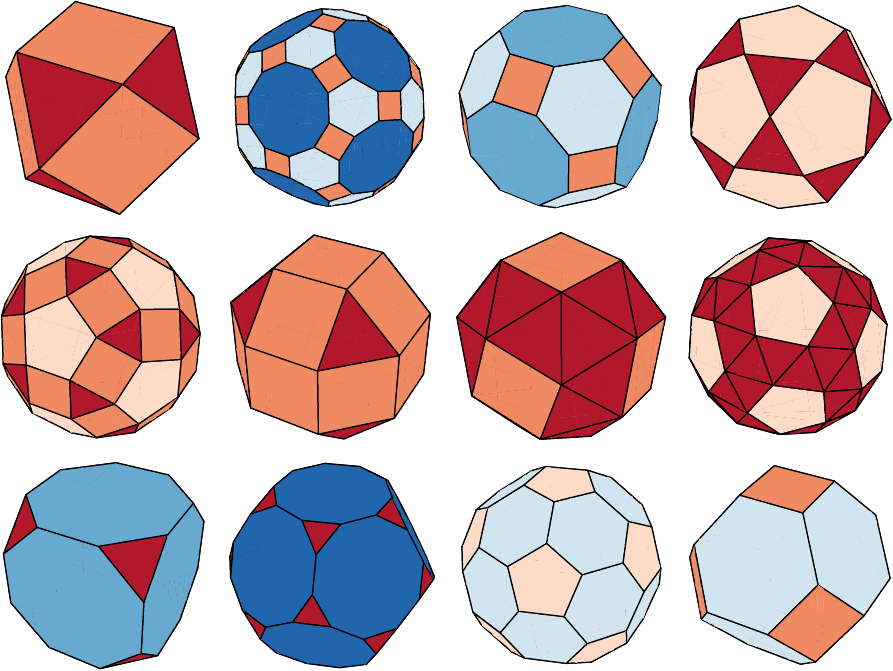}
	\caption{The Archimedean solids, other than the truncated tetrahedron. These shapes can all be circumscribed by a sphere that intersects every one of their vertices. All of the faces of the Archimedean solids are regular polygons, and every vertex of an Archimedean solid is symmetric with every other vertex of the solid (in the sense that they are connected by global isometries). States whose Majorana representations correspond to the vertices of the Archimedean solids are 2-anticoherent, and can be used to attain quantum enhancements in measuring rotations.}
	\label{fig:archimedean}
\end{figure}

Our criteria thus help motivate the Platonic solids as ideal states for measuring rotations. They also point to a much broader class of ideal states. One such extension is the class of Archimedean solids (Figs. \ref{fig:tetrahedrons}-\ref{fig:archimedean}). The 13 Archimedean solids all have discrete rotational symmetries along multiple independent axes. 

For example, the truncated tetrahedron, an Archimedean solid, has the same rotational symmetries as the tetrahedron, with 12 vertices. 
Thus, one can form Majorana constellations made from any combination of $m$ points at each of the vertices a tetrahedron, $n$ points at the vertices of the tetrahedron's dual tetrahedron, $i$ points at the vertices of the associated truncated tetrahedron, and $j$ points at the vertices of the truncated tetrahedron associated with the dual tetrahedron, to obtain a 2-anticoherent state with $N=4n+4m+12i+12j$ quanta (Fig. \ref{fig:tetrahedrons}). Similar constructions can be made with all of the Platonic and Archimedean solids that share rotational symmetries. 

This gives a broad class of states that achieve Heisenberg scaling in estimating the three angles of a rotation. Our symmetry property can be used as a simple geometrical method for generating new 2-anticoherent states.
Such states can be created experimentally in numerous ways: using light's orbital angular momentum degree of freedom to do polarimetry or ellipsometry, as in Ref. \cite{Bouchardetal2017}; likewise, using polarization of light, as in Refs. \cite{Rozema2014thesis,Kimetal2017}; or, using the $2l+1$ orbitals of the $l^\text{th}$ sublevel of a hydrogenlike atom to do magnetometry \cite{steck2003cesium}.
We conjecture that increasing the degeneracy of the Majorana representation of any 2-anticoherent state will yield a new 2-anticoherent state.

\section{Conclusions}
\label{sec:conclusions}
We have presented a thorough investigation of how to attain maximally sensitive measurements of rotations. We derived the quantum Fisher information matrix for Euler angle measurements, and established a quantum Cram\'er-Rao bound on the covariance between the Euler angles being measured. Our bound can always be saturated. 
Specifically, we then focused on the states that optimize the Fisher information: the 2-anticoherent states.
This yielded Heisenberg-scaling precisions in simultaneously measuring all three Euler angles [Eq. (\ref{eq:fisher inverse trace})], offering enhancements over the shot noise limit. 
We showed that the corresponding matrix is always singular for particular combinations of rotation parameters, which is an important consideration in all rotation measurements.
Finally, we mentioned that states whose Majorana representations have two independent rotational symmetries are 2-anticoherent states. This geometrical technique is a powerful way of uncovering new 2-anticoherent states.

Using 2-anticoherent states to optimize rotation measurements has many applications. The quantum enhancements obtained by using 2-anticoherent states can naturally be used in polarimetry and ellipsometry, using light's polarization degree of freedom, and can further be used in precision measurements of electric and magnetic fields, biological samples, and even components of quantum technologies. The exciting field of quantum-enhanced multiparameter estimation has many important ramifications for the near future.

\begin{acknowledgements}
	This work was supported by the NSERC Discovery Award Fund \#480483 and 
	by the Alexander Graham Bell Scholarship \#504825. A.G. acknowledges insightful discussions with H. Ferretti.
\end{acknowledgements}

\begin{appendix}

	\section{Single parameter variances}
	\label{app:single parameter variances}
	In this section we derive a bound on the variances of three rotation parameters estimated using three optimal single-parameter estimation schemes.
	
	We consider as usual the NOON states $\ket{\psi_\text{NOON}}=\frac{\ket{N,0}+\ket{0,N}}{\sqrt{2}}$ (with $N>2$ quanta) and the rotation operators $
			\hat{R}\left(\Phi,\Theta,\Psi\right)=\eu^{-\iu\Phi \hat{S}_z}\eu^{-\iu\Theta \hat{S}_y}\eu^{-\iu\Psi \hat{S}_z}
		$. We try to minimize the variance in estimating $\Phi$, $\Theta$, and $\Psi$ by using three NOON states, each with $N/3$ particles, aligned along various axes. Without loss of generality, we consider the states $\ket{\psi_0(a,b)}=\hat{R}\left(0,a,b\right)\ket{\psi_\text{NOON}}$, where $a=\tan^{-1}(u_y/u_x)$ and $b=\cos^{-1}(u_z)$ parametrize the unit vector $\mathbf{u}=(u_x,u_y,u_z)$ around which the NOON state $\ket{\psi_0(a,b)}$ is most sensitive to measuring rotations.
		
	The state $\ket{\psi_\text{NOON}}$ has \eq{\mathbf{S}=\mathbf{0},\quad\mathcal{S}=\frac{1}{4}\begin{pmatrix}
			N&0&0\\0&N&0\\0&0&N^2
	\end{pmatrix}.} The quantum Fisher information is calculated as before with the $\hat{H}_l$ operators, but now we must take expectation values with respect to the states $\ket{\psi_0(a,b)}$. For this, we use the transformations
\begin{widetext}
	\eq{
		\hat{R}^\dagger\left(0,a,b\right)\hat{H}_\Phi \hat{R}\left(0,a,b\right)&=-\sin a \hat{S}_x+\cos a\hat{S}_z\\
		\hat{R}^\dagger\left(0,a,b\right)\hat{H}_\Theta \hat{R}\left(0,a,b\right)&=\hat{S}_x\cos a\left(\sin b\cos\Phi-\cos b\sin\Phi\right) +\hat{S}_y\left(\cos b\cos\Phi+\sin b\sin\Phi\right) +\hat{S}_z\sin a\left(\sin b\cos\Phi-\cos b\sin\Phi\right)\\
		\hat{R}^\dagger\left(0,a,b\right)\hat{H}_\Psi \hat{R}\left(0,a,b\right)&=\hat{S}_x\left(\sin\Theta\cos\Phi\cos a\cos b+\sin\Theta\sin\Phi\cos a\sin b-\cos\Theta\sin a\right) \\
		+\hat{S}_y&\left(\sin\Theta\sin\Phi\cos b-\sin\Theta\cos\Phi\sin b\right)
		+\hat{S}_z\left(\sin\Theta\cos\Phi\sin a\cos b+\sin\Theta\sin\Phi\sin a\sin b+\cos\Theta\cos a\right).
}
\end{widetext}
We are looking to optimize measurements of a single parameter, using $\text{Var} \left[\theta_i\right]\geq1/4\text{Var}\left[\hat{H}_i\right]$:

	\eq{
		\text{Var}\left[\Phi\right]\geq 
\frac{1}{N^2\left[\mathbf{u}\cdot\mathbf{n}_z\right]^2+N\left[\mathbf{u}\times\mathbf{n}_z\right]^2}\\
		\text{Var}\left[\Theta\right]\geq
		\frac{1}{N^2\left[\mathbf{u}\cdot\mathbf{n}_\Phi\right]^2+N\left[\mathbf{u}\times\mathbf{n}_\Phi\right]^2}\\
		\text{Var}\left[\Psi\right]\geq 
\frac{1}{N^2\left[\mathbf{u}\cdot\mathbf{n}_{\Theta,\Phi}\right]^2+N\left[\mathbf{u}\times\mathbf{n}_{\Theta,\Phi}\right]^2}
,}
for unit vectors 
\eq{
		\mathbf{n}_z=\left(0,0,1\right)\\\mathbf{n}_\Phi=\left(-\sin\Phi,\cos\Phi,0\right)\\\mathbf{n}_{\Theta,\Phi}=\left(\sin\Theta\cos\Phi,\sin\Theta\sin\Phi,\cos\Theta\right)\\
		\mathbf{u}=\left(\sin a\cos b,\sin a \sin b,\cos a\right).}

The variances are similar to those in the known-axis case, in which the denominators look like $N^2\cos^2\theta+N\sin^2\theta$ for a rotation around axis $\left(\sin\theta\cos\phi,\sin\theta\sin\phi,\cos\phi\right)$. 

The unknown axis contributes the parameters $\Theta$ and $\Phi$, and one must choose combinations of $a$ and $b$ that optimize estimates of $\Phi$, $\Theta$, and $\Psi$. The best possible choice for estimating $\Phi$ is by taking $\mathbf{u}=\mathbf{n}_z$. Similarly, the best possible choices for estimating $\Theta$ and $\Psi$ use $\mathbf{u}=\mathbf{n}_\Phi$ and $\mathbf{u}=\mathbf{n}_{\Theta,\Phi}$, respectively; however, these axes cannot be known a priori.

Each scheme can only use $N/3$ particles, so the optimal combination of these three variances for various choices of $a$ and $b$ yields Eq. (\ref{eq:single parameter variances}) above, where we have chosen $\mathbf{n}_\Phi\cdot\mathbf{u}=\cos\theta_1$ and $\mathbf{n}_{\Theta,\Phi}\cdot\mathbf{u}=\cos\theta_2$. 
However, this idea of combining the single-parameter Cram\'er-Rao bounds for a multiparameter estimation technique, like in Refs. \cite{Humphreysetal2013,BaumgratzDatta2016}, cannot be sufficient. If it were the case, we could simply take the diagonal components of Eq. (\ref{eq:QFIM}), invert them, and achieve Heisenberg scaling precisions for $\Phi$, $\Theta$, and $\Psi$ regardless of rotation angle, which is impossible. The difference here is that we cannot treat two out of the three parameters as known while estimating the third, so one cannot truly subdivide the system and estimate a single parameter for each section while being ignorant of the other parameters. Only a true multiparameter estimation technique can succeed in our case.

\section{Divergences in every rotation angle parametrization}
\label{app:divergences}
We here discuss the fact that every rotation angle parametrization yields singular quantum Fisher information matrices for particular combinations of rotation angles. 

We start by choosing three rotation parameters $a$, $b$, and $c$. Any triplet of rotation parameters can in principle be obtained from any other such triplet (e.g., by equating the rotation matrices for the various parametrizations and solving the resulting nonlinear equations), so we use our parametrization $
\hat{R}\left(\Phi,\Theta,\Psi\right)=\eu^{-\iu\Phi \hat{S}_z}\eu^{-\iu\Theta \hat{S}_y}\eu^{-\iu\Psi \hat{S}_z}
$ for variables \eq{\Phi=\Phi(a,b,c),\quad\Theta=\Theta(a,b,c),\quad \Psi=\Psi(a,b,c).}

Next, we formally compute the derivatives
\eq{
d_k\hat{R}=-\iu\left(\hat{H}_\Phi d_k\Phi+\hat{H}_\Omega d_k\Omega+\hat{H}_\Psi d_k\Psi\right)\equiv-\iu\hat{H}_k\hat{R},
}
for $k\in\left(a,b,c\right)$ and $d_k\equiv d/dk$. In the $\boldsymbol{\tilde{\theta}}=(a,b,c)$ parametrization, the quantum Fisher information matrix has components related to the $\boldsymbol{\theta}=\left(\Phi,\Theta,\Psi\right)$ QFIM:
\eq{
\left[\mathbf{I}_{\boldsymbol{\tilde{\theta}}}\right]_{i,j}
=\begin{pmatrix}
	d_i\Phi&d_i\Theta&d_i\Psi
\end{pmatrix}
\mathbf{I}_{\boldsymbol{\theta}}
\begin{pmatrix}
	d_j\Phi\\d_j\Theta\\d_j\Psi
\end{pmatrix}.
}

We recognize the Jacobian 
\eq{
\mathbf{J}=
\begin{pmatrix}
	d_a\Phi&d_b\Phi&d_c\Phi\\d_a\Theta&d_b\Theta&d_c\Theta\\
	d_a\Psi&d_b\Psi&d_c\Psi\\
\end{pmatrix}
}
and the transformation $\mathbf{I}_{\boldsymbol{\tilde{\theta}}}=\mathbf{J}^\text{T}\mathbf{I}_{\boldsymbol{\theta}}\mathbf{J}$. We have already shown that the matrix $\mathbf{I}_{\boldsymbol{\theta}}$ is singular at angle $\Theta=0$, because $\text{Det}\left[\mathbf{I}_{\boldsymbol{\theta}}\right]\propto\sin^2\Theta$. The new matrix $\mathbf{I}_{\boldsymbol{\tilde{\theta}}}$ is singular whenever $\mathbf{I}_{\boldsymbol{\theta}}$ is singular, unless $\text{Det}\left[\mathbf{J}\right]$ diverges as $1/\sin\Theta$. This singularity is a coordinate singularity; there is no set of coordinates that can cover a sphere without such singularities. The best that can be done is to hope for a parametrization whose coordinate singularity occurs at a different set of coordinates.

The only possibility of $\mathbf{I}_{\boldsymbol{\tilde{\theta}}}$ being invertible at $\Theta=0$ is if $\text{Det}\left[\mathbf{J}\right]\propto 1/\sin\Theta$ for all values of $\Phi$ and $\Psi$ at that point. Namely, one would need $d_a\Theta\left( d_c\Phi d_b\Psi- d_b\Phi d_c\Psi\right)+ d_b\Theta\left( d_c\Phi d_a\Psi- d_a\Phi d_c\Psi\right)+ d_c\Theta\left( d_b\Phi d_a\Psi- d_a\Phi d_b\Psi\right)\propto1/\sin\Theta$ for all values of $\Phi$ and $\Psi$ as $\Theta\to 0$. This requires that $d_a\Theta\propto d_b\Theta\propto d_c\Theta\propto 1/\sin\Theta$ as $\Theta\to0$. If that were the case, then one would never be able to estimate the value of $\Theta$ from the three parameters $a$, $b$, and $c$ near $\Theta=0$. 

Still, if one indeed had $\text{Det}\left[\mathbf{J}\right]\propto 1/\sin\Theta$, we would expect $\text{Det}\left[\mathbf{J}\right]\to0$ at a different value of $\boldsymbol{\theta}$, 
because coordinate singularities are present regardless of parametrization. 
Since the rotation parameters are unknown a priori, it is impossible to definitively choose a parametrization that is guaranteed to be nonsingular for a given rotation measurement. Perhaps one could avoid the divergences by using $N/2$ particles in each of two separate rotation measurements whose singular coordinates do not coincide.

\end{appendix}

%

\end{document}